\documentclass[longbibliography,final,aps,prx,amsmath,amssymb,superscriptaddress,twocolumn]{revtex4-2}

\usepackage[T1]{fontenc}
\usepackage{lmodern}
\usepackage{amsfonts}
\usepackage{amsmath,fixmath}
\usepackage{microtype}
\usepackage[usenames,dvipsnames]{xcolor}
\definecolor{shadecolor}{gray}{0.9}
\usepackage{graphicx}
\usepackage{overpic}
\usepackage{booktabs}
\usepackage{tabularx}
\usepackage{algorithm}
\usepackage{algpseudocode}
\usepackage{natbib}
\usepackage[colorlinks,linktoc=all]{hyperref}
\usepackage[all]{hypcap}
\hypersetup{citecolor=Violet}
\hypersetup{linkcolor=black}
\hypersetup{urlcolor=MidnightBlue}
\usepackage[nameinlink]{cleveref}

\newcommand{\be}{\begin{equation}}
\newcommand{\ee}{\end{equation}}
\newcommand{\bea}{\begin{eqnarray}}
\newcommand{\eea}{\end{eqnarray}}
\newcommand{\mbeq}{\stackrel{!}{=}} 
\newcommand{\ra}{\rightarrow}
\newcommand{\f}[2]{\frac{#1}{#2}}

\newcommand{\ccup}[1]{\left\{#1\right\}}

\newcommand{\bup}[1]{\left(#1\right)}

\DeclareMathOperator*{\argmin}{arg\,min}

\newcommand{\Graph}{\mathcal{G}}
\newcommand{\Nodes}{\mathcal{V}}
\newcommand{\Edges}{\mathcal{E}}

\newcommand{\Ltwonorm}[1]{\lvert| {#1} \rvert|_2}
\newcommand{\Flux}{F}
\newcommand{\Iedge}{e}
\newcommand{\Icomm}{i}

\newcommand{\NComm}{M}

\newcommand{\CommSet}{1,\ldots,\NComm}

\renewcommand{\ref}[1]{[\ref{#1}]}

\usepackage[caption=false]{subfig}

\usepackage[textsize=tiny,color=SkyBlue]{todonotes}

\usepackage[normalem]{ulem} 

\newcommand{\kl}{Kirchhoff's}

\begin{document}

\title{Designing optimal networks for multi-commodity transport problem}

\author{Alessandro Lonardi}
\email{alessandro.lonardi@tuebingen.mpg.de}
\affiliation{Max Planck Institute for Intelligent Systems, Cyber Valley, T{\"u}bingen 72076, Germany}
\author{Enrico Facca}
\affiliation{Centro di Ricerca Matematica Ennio De Giorgi, Scuola Normale Superiore, Piazza dei Cavalieri, 3, Pisa, Italy}
\author{Mario Putti}
\affiliation{Department of Mathematics ``Tullio Levi-Civita'', University of Padua, Via Trieste 63, Padua, Italy}
\author{Caterina De Bacco}
\email{caterina.debacco@tuebingen.mpg.de}
\affiliation{Max Planck Institute for Intelligent Systems, Cyber Valley, T{\"u}bingen 72076, Germany}

\begin{abstract}
Designing and optimizing different flows in networks is a relevant problem in many contexts. While a number of methods have been proposed in the physics and optimal transport literature for the one-commodity case, we lack similar results for the multi-commodity scenario. In this paper
    we present a model based on optimal transport theory for finding optimal multi-commodity flow configurations on networks. This model
       introduces a dynamics that
      regulates the edge conductivities to
      achieve, at infinite times, a minimum of a Lyapunov functional given by the sum of a
      convex transport cost and a concave infrastructure cost.
    We show that the long time asymptotics of this dynamics are the solutions of a standard constrained optimization problem that generalizes the one-commodity framework. Our results provide new insights into the nature and properties of optimal network topologies. In particular, they show that loops can arise as a consequence of distinguishing different flow types, complementing previous results where loops, in the one-commodity case, were obtained as a consequence of imposing dynamical rules to the sources and sinks or when enforcing robustness to damage.  Finally, we provide an efficient implementation of our model which convergences faster than standard optimization methods based on gradient descent.
\end{abstract}

\maketitle

\section{Introduction}
Optimizing networks for the distribution of quantities like passengers
in a transportation network or data packets in a communication network is a relevant matter for network planners.  Similar problems arise in natural systems like river basins and vascular networks. A variety of models have been proposed to study these systems within an optimization framework  \cite{maritan,bohn2007structure,corson2010fluctuations,sinclair1996}. The standard goal is to find the values of flow and the network topology that minimize a transportation cost. A common choice for this cost is the total power dissipation  \cite{bohn2007structure,maritan,hu2013adaptation,ronellenfitsch2016global,katifori2010damage, ronellenfitsch2019phenotypes, kaiser2020discontinuous}, but alternatives can be adopted depending on the application, see for instance \cite{kirkegaard2020optimal}. More recently, different approaches based on a dynamical adaptation of network properties coupled with conservation laws have been proposed \cite{ronellenfitsch2016global,hu2013adaptation}. These models can be reformulated within the framework  of  optimal transport theory, following the work of \cite{bonifaci2012physarum, santambrogio2007optimal,facca2016towards,facca2019numerics,facca2020branch,baptista2020network,bonifaci2020}.   Very efficient computational techniques have been developed for solving such optimal transport based models \cite{facca2016towards,facca2019numerics,facca2020branch}.

In all these systems there is a unique undistinguishable flow traveling through the network. However, it may occur that flows of different types compete in the network infrastructure,
yet all the physical models mentioned above have been developed for one type of flow only. One could use these methods to analyze multi-commodity problems by either aggregating together all flow types or by treating them independently. In either case, one loses the important information of how interacting commodities affect the flow, which constitutes the multi-commodity character of these settings.
Multi-commodity-specific methods that rely on standard optimization suffer of high computational costs caused by the simultaneous assignment of multiple interacting paths to minimize a global cost function. As a consequence, existing multi-commodity flow algorithms rely on ignoring these interactions, or use greedy heuristics and approximations that lead to sub-optimal solutions \cite{salimifard2020multicommodity}. Approaches based on statistical physics and message-passing algorithms have improved results \cite{yeung2013physics,yeung2013networking} but remain computationally costly.

In this
  work, we propose a model to design the topology of optimal
  networks where multiple resources are moved together. This is based on
  principles of optimal transport theory similar to those studied in \cite{bonifaci2020, baptista2020network}. Assuming
  potential-driven flows, this optimal design problem is posed as
that of finding the distribution of multi-commodity fluxes
  that minimize a global cost functional, or equivalently, as that of finding the optimal edge conductivities.
The cost functional is the multi-commodity extension of the optimal-transport Lyapunov functional
  proposed in~\cite{facca2019numerics,facca2020branch}. It is given by
  the sum of the convex cost incurred in transporting all the
  commodities across the network, summed to a concave cost proportional to
  the total flux on the network. This second term can be
    interpreted as the cost for building and maintaining the transport
    infrastructure, and controls traffic congestion on the network edges by either distributing fluxes on many edges, or by concentrating them on fewer edges following a principle of economy of scale.

Additionally, we show that the problem of minimizing the
  proposed cost functional is equivalent to a constrained optimization
  problem that generalizes the one-commodity case. The optimal
  distribution of fluxes is used to identify the optimal network topology by discarding edges where conductivities are small.  Within this optimization framework, numerical experiments supported by analytical evidence lead to the important result that optimal network topologies may have loops as a consequence of distinguishing flow types.
Generally, loops are pervasive in both natural and anthropic networks \cite{nelson1997leaf,schaffer2006two,nardini2012trade,katifori2010damage,banavar1999size}.
However, in one-commodity settings, several studies have shown that trees are often optimal \cite{maritan,bohn2007structure}, while few results show that loops can be obtained by fluctuating flows or by aiming at increased robustness to damage \cite{corson2010fluctuations, hu2013adaptation,katifori2010damage}. This implies either changing the type of cost function or introducing stochasticity in the sources and sinks. Instead, in our multi-commodity model loops
  emerge naturally as a consequence of the presence of different flow
  types.

In order to minimize the
  highly non-linear and non-convex cost functional mentioned before, we
  propose a particular set of dynamical equations for the edge
  conductivities, generalizing to a multi-commodity scenario those proposed in~\cite{bonifaci2020,
    baptista2020network}, and find their stationary solution. We demonstrate that the cost functional is indeed a Lyapunov functional
  (i.e. it is strictly decreasing along the solution trajectories) for
  the proposed dynamics.  Altogether, our results extend the
  theoretical insights of two separate lines of literature, optimal
  transport and network dynamics.
Two principled
  algorithms for solving the multi-commodity problem are proposed.
  They have similar computational complexity that largely improves that of techniques based on gradient descent or Monte Carlo methods, thus making the model scalable to large datasets and the only computationally viable optimization alternative for large problems.

\section{Model}
 Consider a graph $\mathcal{G}$ made of a set  of
  $N$ nodes $\mathcal{V}$ interconnected by a set
  $\mathcal{E}$ of $E$ edges. We want to model transport
  through the network of $M\geq 1$ commodities, each identified by a
  color.  The inflow/outflow rate of each commodity is given by a vector
  $S^{i}\in \mathbb{R}^{N}$ such that $\sum_{v} S^{i}_{v}=0$ for all
  $i = \CommSet$ to ensure global mass preservation. Let the ``colored-flux'' $F_{e}=(F_{e}^{1},\dots,F_{e}^{M})$ be a
  vector with entries $F_{e}^{i}$, which represent the commodities
  flux passing through edge $e$. In standard one-commodity cases, the
  flux per unit time could represent a water or an electrical
  current, and typically is ``colorless'', i.e. $F_{e}$ is a scalar
  quantity. In turn, the components $F_e^i$ can be thought as fluxes of immiscible substances traveling through the same edge.
  Denote with $B$ the signed network incidence matrix, with
  entries $B_{ve}=+1,-1$ if node $v\in \mathcal{V}$ is the starting or
  ending point of edge $e\in \mathcal{E}$, respectively, and zero
  otherwise. We require the flux to obey the ``colored'' local
  Kirchhoff's law: \be \label{eqn:kirkoff} \sum_{e \in \mathcal{E}}
  B_{ve}\,F_{e}^{i} =S_{v}^{i} \quad\,
  \forall v \in \mathcal{V}, \, \forall i = \CommSet \quad, \ee where each edge $e=(u,v)$ has
  length $\ell_{e} > 0$. We could assume that the components (colors) of the
  flux derive from differences of a colored-potential (pressure)
  defined on nodes $p_{v}^{i}$, and a colored-conductivity
  $\mu_{e}^{i}$: \be\label{eqn:flow} F_{e}^{i}
  =\f{\mu_{e}^{i}}{\ell_{e}} \, \bup{p^{i}_{u}-p_{v}^{i}} \quad.  \ee

\begin{figure}[htp]
\begin{center}
\includegraphics[width=1.0\columnwidth,trim={0.2cm 0.1cm 0.22cm 0.1cm },clip]{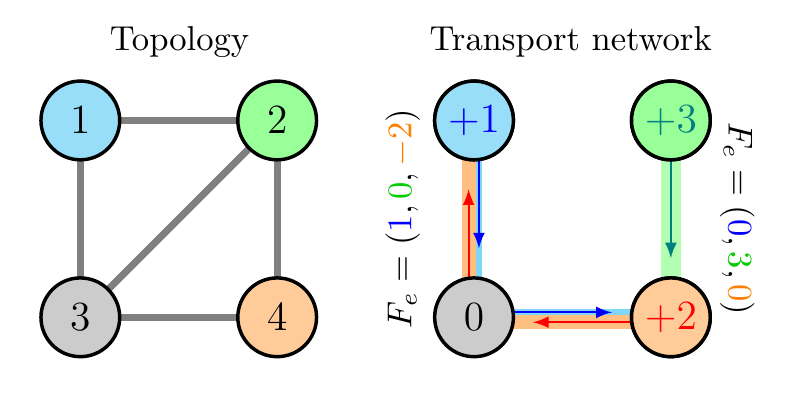}
\caption{Multi-commodity problem illustration for $\NComm=3$. (Left) Topology of the graph, numbers inside nodes correspond to their indexes $v$. (Right) An admissible configuration of fluxes.  Here, number inside nodes correspond to their mass inflow, we assume for each commodity to have its mass concentrated in a single vertex; in the gray node no mass is entering nor exiting, i.e. $S_3^i = 0$ for every $i$. Widths of edges are drawn proportional to $F_e^i$, in case all $F_e^i =0$ links are not drawn, and arrows denote the direction of the colored fluxes $F_e^i$. Notice how blue and orange mass share the same edges, thus creating possible traffic congestion. The inflow/outflow rates of each color are are $S^{1}=(1,0,0,-1)$ (blue), $S^2=(0,3,0,-3)$ (green) and $S^3=(-2,0,0,2)$ (orange).}
\label{fig:example}
\end{center}
\end{figure}

The commodity index $i$ can be any arbitrary attribute of the mass traveling through the network without impacting the validity of our model. In fact, the important idea behind multi-commodity optimization is that mass of different type interacts while being transported in a shared infrastructure, and a suitable cost needs to be minimized. In Fig.~\ref{fig:example} we show a simple example of the model construction.

Up to this point, we have a set of independent one-commodity flows,
one per color $i$. Taking them separately and then superimposing each
individual flux or conductivity a posteriori would be a naive
strategy, neglecting possible complex interactions.  For example,
  it may be more convenient to gather multiple flows through one
  channel with high capacity (the conductivity $\mu_e^i$). More generally, the optimal network design mechanism must take into account all commodities
  at once.
  Deciding how this should be done is an open problem in the
context of optimal transport theory, the approach we take here.

\subsection{Introducing a shared conductivity}\label{ssec:mu}
Our first model assumption is that
all the conductivities must be equal, namely:
\be\label{eqn:constr}
\mu_{e}^{i}\mbeq \hat{\mu}_{e} \quad \quad \forall e \in \mathcal{E},\, \forall i = \CommSet \quad.
\ee
The quantity $\hat{\mu}_{e} $ plays the role of a ``colorless'' conductivity. Given that the conductivity can be seen as proportional to the size of an edge, Eq. (\ref{eqn:constr}) can be interpreted as allocating the same edge capacity for the different colors.
This is a reasonable assumption in systems for which there is no priority between commodity types or users. In communication networks like the internet, this captures the situation where all users share the same bandwidth, and no privileged user exists who has access to more bandwidth, which is often the case.
Notice, however, that the flux $F_{e}^{i}$ still depends on the color, because the difference in potential does.  This implies that users can transfer different amounts of data packages, with potential for traffic congestion when they overload the network, e.g. when streaming videos.
This is one of the many possible alternatives of coupling between colors.  Other more complex choices could be made, for instance by introducing an explicit coupling involving the fluxes, or opting for controlling some global functions of the conductivities, e.g. their sum or the magnitude of their fluctuations across colors. However, we find that our choice,  while being analytically convenient, it allows for a rigorous generalization of the one-commodity case with fixed and fluctuating loads \cite{bohn2007structure, maritan,katifori2010damage,corson2010fluctuations,hu2013adaptation} and leads to rich topological behaviors, as we show below.

\subsection{The dynamics}
Having defined how colors move through the network, we now turn our attention on describing the mechanism to design the network. Formally, we propose the following dynamics for the colorless conductivity:
\be\label{eqn:dyn}
\dot{\hat{\mu}}_{e}(t) =   \hat{\mu}_{e}^{\beta}(t) \f{||\Delta P_{uv}||^{2}_{2}}{\ell_e^2}\, - \hat{\mu}_{e}(t)\quad \forall e=(u,v) \in \mathcal{E}\quad,
\ee
where we define $\Delta P_{uv}$ as a vector of pressure differences with entries $\Delta P_{uv}^{i}=p^{i}_{u}-p^{i}_{v}$ and $||\Delta P_{uv}||^{2}_{2}=\sum_{i=1}^{M} \bup{p^{i}_{u}-p^{i}_{v}}^{2}$.  Note that $p^{i}$, and thus $\Delta P_{uv}$, are implicit functions of $\hat{\mu}(t)$ because of Eq.~(\ref{eqn:kirkoff}) and~(\ref{eqn:flow}).

The parameter $\beta$ determines the type of optimization associated to this dynamics. In the standard one-commodity case, for $\beta<1$ one aims at minimizing traffic congestion and obtains loopy topologies, for $\beta>1$ the aim is to consolidate paths and optimal networks are trees; the case $\beta=1$ is shortest path-like. This dynamics describes a feedback mechanism. If the total flux through an edge is large, its conductivity increases. If the flux decreases, the conductivity decreases over time and becomes negligible when no flux occurs.
The systems of Eq.~(\ref{eqn:kirkoff})-(\ref{eqn:dyn})
represents our model for multi-commodity flow optimization.

In the presence of only one commodity, our model is similar to the dynamics used to solve the basis pursuit problem on networks \cite{facca2020branch} and as a principled mechanism for filtering networks from redundancies \cite{baptista2020network}. However, both cases are limited to one-commodity scenarios. A similar dynamics is also proposed in \cite{hu2013adaptation}, where the authors focus on the average time-evolution of a stochastic model with fluctuating loads. Analogously to these one-commodity cases, one can efficiently solve the system in Eq.~(\ref{eqn:kirkoff})-(\ref{eqn:dyn}) using optimized numerical methods, however in our case the complexity increases with the number of colors, see Appendix Sec. \ref{secAPX:compcost} for more details.
Our model also bears a close mathematical relationship to a recent work, where similar ideas have been studied in a multi-commodity setup \cite{bonifaci2020}. Beyond the fact that this work focuses on the case $\beta=1$ (i.e. shortest path-like) and thus on a convex optimization scenario, there is one other main conceptual difference with our model. Notice that Eq.~(\ref{eqn:dyn}) couples together the various colors by means of $f(\Delta P_{uv})=||\Delta P_{uv}||_{2}^{2}$, i.e. the 2-norm squared of the pressure difference. Instead, they consider the 1-norm and 2-norm (not squared). Analyzing the solutions of the dynamics under different $f(\Delta P_{uv})$ is an interesting avenue for future work.

The key insight of optimal transport theory, is that
Eq. (\ref{eqn:dyn}) admits a a Lyapunov functional (a functional
decreasing in time along solution trajectories) having the nice interpretation of being the transportation cost:
\be
\label{eqn:lyap}
\displaystyle\mathcal{L}_{\beta}(\{\hat{\mu}_e\}) =  \f{1}{2}\sum_{i,v} p^i_v(\{\hat{\mu}_e\})S^i_v  + \frac{\sum_{e}\ell_{e}\hat{\mu}_{e}^{2-\beta} }{2(2-\beta)} \quad,
\ee
where $p^i_v(\{\hat{\mu}_e\})$ is a function implicitly defined as solution of Eq.~(\ref{eqn:kirkoff})-(\ref{eqn:flow}) when imposing Eq.~(\ref{eqn:constr}). The first term corresponds to the energy dissipated during transport, it can be interpreted as the operating costs, whereas the second is the cost of designing the infrastructure. The equilibrium point of $\hat{\mu}_{e}$ is stationary at the previous Lyapunov functional, and for $\beta\leq1$ it acts also as the global minimizer due to its convexity.  For $\beta>1$, while the first term (operating cost) is convex, the second  (infrastructural cost) is not. As a consequence, the transportation cost is not convex, thus in general the functional will present a rich landscape with several local minima towards which the dynamics will be attracted.

We formally show that Eq.~(\ref{eqn:lyap}) defines a well-defined Lyapunov functional for the dynamics of Eq.~(\ref{eqn:dyn}) in Appendix \ref{secAPX:lyapunov_dyn}, following similar arguments as in \cite{bonifaci2020}.
This extends the work of Bonifaci et. al. \citep{bonifaci2012physarum}, where a similar functional has been proposed to complete the characterization of the dynamics regulating slime molds' evolutionary feedback mechanism.

\subsection{Mapping to standard optimization setups}
Although not evident, our dynamics is connected with an optimization problem analogous to previous models for the one-commodity case \cite{bohn2007structure,maritan}.
 Specifically, the stationary solutions of our system minimizes the network total transportation cost
 $J=\frac{1}{2}\sum_{e\in \mathcal{E}} \,\f{\ell_{e}}{ \hat{\mu}_{e}}\,||F_{e}||^{2}_{2}$ subject to the global constraint of constant material cost $\sum_{e \in \Edges}\ell_{e}\,\hat{\mu}_{e}^{2-\beta}=K^{2-\beta}$ and local Kirchhoff's law on nodes as in Eq. (\ref{eqn:kirkoff}) (using Kirchhoff's law one can show that $J$ is equivalent to the first term in Eq. (\ref{eqn:lyap}), for more details see Appendix \ref{secAPX:lyap_lagrange}).
Formally the optimization problem is:
 \be\label{eqn:opt1}
\ccup{\hat{\mu}^{*}_{e}},\ccup{F^{*}_{e}}= \argmin_{\ccup{\hat{\mu}_{e}},\ccup{F_{e}}} \ccup{\frac{1}{2}\sum_{e\in \mathcal{E}} \,\f{\ell_{e}}{ \hat{\mu}_{e}}\,||F_{e}||^{2}_{2}} \quad,
 \ee
 such that
 \bea
 \sum_{e\in \mathcal{E}}\ell_{e}\,\hat{\mu}_{e}^{2-\beta}&=&K^{2-\beta}\quad,\label{eqn:opt2}\\
 \sum_{e \in \mathcal{E}} B_{ve}\,F_{e}^{i} &=&S_{v}^{i} \quad\, \forall v \in \mathcal{V}, \, \forall i = \CommSet \quad.\label{eqn:opt3}
 \eea
 This optimization problem is analogous to that in \cite{bohn2007structure}, except here the flux appears in terms of its 2-norm. As in the one-commodity case, this leads to an optimal configuration where the conductivities similarly scale with the fluxes,
 \be\label{eqn:scaling}
  \hat{\mu}_{e}\sim ||F_{e}||_{2}^{2/(3-\beta)} \quad,
  \ee
and the proportionality constant can be fully determined analytically (see Appendix \ref{secAPX:dyn2opt} for detailed derivations). Using Eq. (\ref{eqn:scaling}) we can rewrite the total transportation cost in terms of the flux as
  \be
  \label{eqn:cost}
  J_{\Gamma}=\sum_{e\in \mathcal{E}} \,\ell_{e}\,||F_{e}||_{2}^{\Gamma} \quad,
  \ee
 where $\Gamma=2\,(2-\beta)/(3-\beta)$, which is analogous to the optimization problem of Banavar et al. \cite{maritan}, where there was no conductivity in the setup.
Notice that all these results generalize the one-commodity case \cite{bohn2007structure,maritan} by means of the 2-norm $||F_{e}||_{2}$ of the colored flux. If there were only one color, and thus $||F_{e}||^{2}_{2}= F_{e}^{2}$, our model reduces exactly to them. Similar relations can be obtained with a stochastic approach as the one proposed in \cite{corson2010fluctuations,katifori2010damage, ronellenfitsch2019phenotypes}, but by considering ensemble averages instead of the 2-norm of the fluxes. In these works, the authors study a setup where sources and sinks' positions are extracted randomly from a distribution on the network nodes. They also find loops in non-trivial regimes. While the mathematical formulations show some similarities, there are main conceptual differences between these models and ours. These approaches are stochastic, thus the main quantities are calculated with ensemble averages and loops arise as a consequence of stochastic fluctuations, or when randomly cutting edges in the network. Instead, our problem is deterministic and loops arise as a result of an optimization process while assuming a shared conductivity.

Solving this optimization problem directly by means of gradient descent is computationally expensive (see Appendix Sec.~\ref{secAPX:algorithms}). Methods relying on Monte Carlo schemes \cite{bohn2007structure} can also be computationally demanding, and they are valid only when the optimal topology is known to be a tree. Instead, we derive update rules which have similar complexity as that of finding the steady states of our dynamics, and can be implemented with efficient numerical solvers. They consist in iterating between updating conductivities and fluxes as follows:
\bea
\hat{\mu}_{e} &=& \f{||F_{e}||^{2/(3-\beta)}_{2}}{ \bup{ \sum_{e}\ell_{e}\,||F_{e}||^{2(2-\beta)/(3-\beta)}_{2}}^{1/(2-\beta)}}\, K\quad,\label{eqn:iterativeMu}\\
F_{e}^{i}  &=& \f{\hat{\mu}_{e}}{\ell_{e}}\,\bup{p_{u}^{i}-p_{v}^{i}}\label{eqn:iterativeF} \quad,
\eea
complemented with Kirchhoff's law in Eq. (\ref{eqn:kirkoff}), and can be put within the framework of fixed-point iterations. This generalizes results obtained adopting a similar approach for the one-commodity case \cite{bohn2007structure,corson2010fluctuations}.
We make available an open source implementation of the two approaches which we summarize here: finding the steady state of the dynamics by solving the systems of Eq.~(\ref{eqn:kirkoff})-(\ref{eqn:dyn})  (Dynamics); extracting the solution of the optimization problem with the iterative updates of Eq.~(\ref{eqn:iterativeMu})-(\ref{eqn:iterativeF}) (Optimization).  We provide a pseudo-code for each of these in Appendix Algorithms (\ref{dynamics_algAPX}) and (\ref{optimization_algAPX}).
They have similar computational complexity that scales as $\mathcal{O}(M\,N^{2})$, and are much faster than techniques based on gradient descent, see Appendix Sec.~\ref{secAPX:algorithms}.

\section{Analysis of the optimal topologies}\label{sec:toplogies}

\subsection{Optimal topologies may have loops}\label{sec:loops}
Now, we address the important question of which network topologies are optimal for the cost in Eq. (\ref{eqn:cost}). For the analogous models in the one-commodity case, there is a phase transition at $\beta=\Gamma=1$ where optimal networks pass from being trees ($1< \beta< 2$, $0<\Gamma<1$) to containing loops ($0< \beta<1$, $1<\Gamma<4/3$) \cite{bohn2007structure,maritan}, see \cite{kaiser2020discontinuous} for a thorough investigation of this transition. Remarkably, we obtain that in the multi-commodity case, loopy structures can be optimal also in the regime where trees were optimal in the previous models, depending on the values and locations of sources $S_{v}^{i}$ and on the edge lengths $\ell_{e}$.

The loopy structures in what was previously a tree-like regime arise from the colored Kirchhoff's law (\ref{eqn:kirkoff}), distinguishing different commodities entering and exiting a node. Had we imposed a similar but ``colorless'' constraint $\sum_{i} \sum_{e} B_{ve} F_{e}^{i} =\sum_{i} S_{v}^{i} $, trees would have been optimal.

\begin{figure}[htp!]
\begin{center}
\includegraphics[scale=0.78,trim={0cm 0.2cm 0cm 0.5cm},clip]{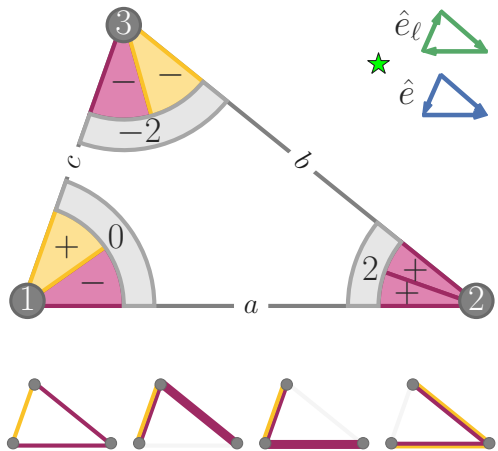}
\caption{\label{fig:tree_toymodel_panel} Toy model where loops are optimal.  Here $M=2$, hence only two colors move though the network. The triangle network has source vectors $S^{1}=(+1,-1,0),\,S^{2}=(-1,+2,-1)$, $\ell_{a}=\ell_{b}=1.5 ,\, \ell_{c}=1$. The grey patches denote the net loads of each node when ignoring the colors. On the bottom we show one loopy solution on the left and three trees on the right. The green and blue arrows denote the orientation of the loop defined in Appendix \ref{secAPX:trimxia} and of the edges, respectively.  In detail, the loop has fluxes $F_{a}=(0,-1),\,F_{b}=(0,-1),\, F_{c}=(-1,0)$, the leftmost tree has $F_{a}=(0,0), \, F_{b}=(0,-2),\, F_{c}=(-1,+1)$  (similarly for the other two). The green star refers to the topology of the toy model used in Fig.~\ref{fig:phaseD}.
}
\end{center}
\end{figure}

\begin{figure*}[htp]
\begin{center}
\includegraphics[width=1.7\columnwidth,trim={0.4cm 0.0cm 1.8cm 0.0cm},clip]{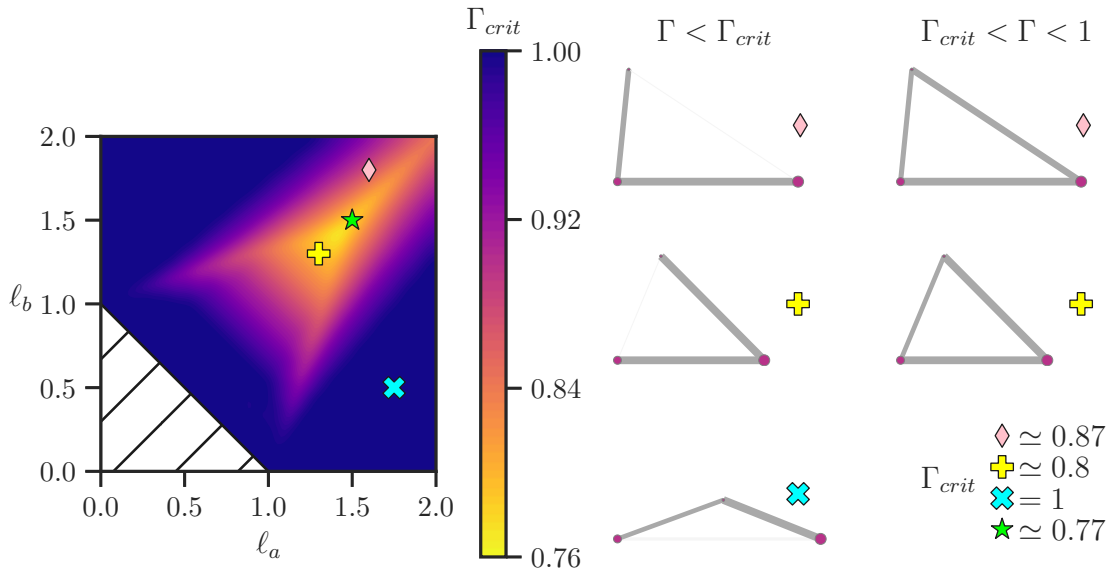}
\caption{\label{fig:phaseD} Phase diagram in $\ell$. $\Gamma_{crit}$ denotes the minimum value of $\Gamma$ above which loops are optimal. The setup is the same as for the toy model in Fig. \ref{fig:tree_toymodel_panel}. Values of $\Gamma_{crit}$ are found by solving Eq. (\ref{eq:mincost_toymodel}); $\ell_{c}=1$. The area under the triangular surface in white background is not allowed as the triangular geometry is not defined there. One can notice that there is an entire region where $\Gamma_{crit}\leq 1$, inside it loops are optimal. On the r.h.s of the Figure we show three different topologies, i.e. choices of $\ell_a, \ell_b$ for which optimal solutions can be loopy; these are associated with the markers drawn on the heatmap, the green star is the configuration of the toymodel in Fig. \ref{fig:tree_toymodel_panel}. In particular, running our dynamics on the pink-diamond graph (resp. the yellow-plus) leads to a loopy configuration for $\Gamma > \Gamma_{crit}$ or to a tree if $\Gamma < \Gamma_{crit}$. Running our dynamics on the blue-cross graph returns always since its $\Gamma
_{crit}$ is equal to $1$. Widths of the edges are proportional to the final $||F_{e}||_2$, not visible edges have negligible fluxes; nodes' dimensions are proportional to their the inflowing mass. In the bottom right portion of the panel we report the values of $\Gamma_{crit}$ obtained solving Eq. (\ref{eq:mincost_toymodel}) fixing $\ell_a, \ell_b$ as given by the markers. }
\end{center}
\end{figure*}

\begin{figure*}[htb]
\begin{center}
\begin{overpic}[width=2.2\columnwidth,trim={1cm 0 0 0},clip]{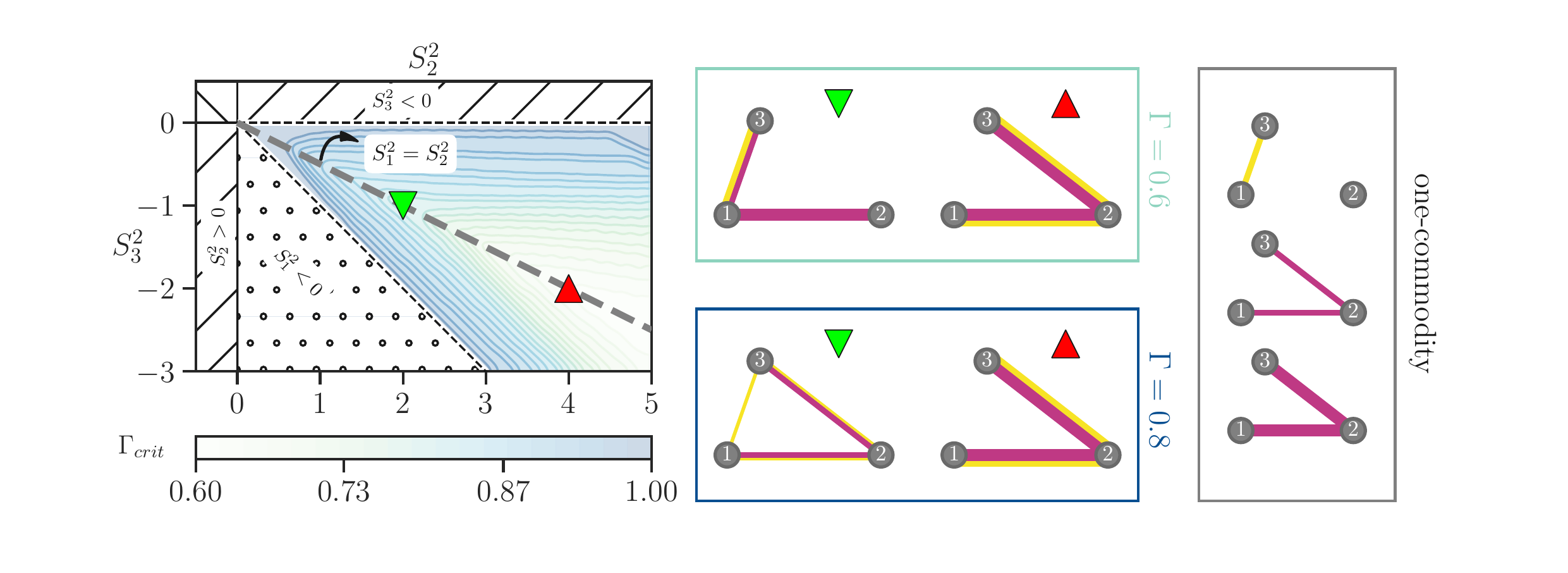}
\put (23,1) {(a)}
\put (55,1) {(b)}
\put (81,1) {(c)}
\end{overpic}
\caption{\label{fig:panel_interaction_fluxes}  Interaction between commodities.  The figure shows how changing the load of one commodity influences the path taken by others. (a) We plot the heat map of $\Gamma_{crit}$ obtained using Eq.~(\ref{eq:mincost_toymodel}), with $S_1^1 = - S_3^1 = +1$ and for different configurations of the purple commodity ($i=2$), i) $S_2^2=+2$, $S_3^2 = S_1^2 = -1$; ii) $S_2^2=+4$, $S_3^2 = S_1^2 = -2$. The areas under the white surface correspond to regions where node $1$ is a sink, resp. nodes $2$ and $3$ are sources, for the purple mass. The green and the red triangular markers denote the configurations we discuss in Sec \ref{ssed:interaction_colors} and in the rest of the panel.  (b) Optimal graphs for the green and the red triangle, fixing $\Gamma = 0.6$ and $\Gamma = 0.8$. The width of each edge $e$ is proportional to $|\Flux_\Iedge^\Icomm|$ for each color. (c) One-commodity solutions obtained injecting only the yellow or the purple mass in the triangle network.}
\end{center}
\end{figure*}

\subsection{Phase diagram tree--loops ($S$ fixed)}\label{sec:phaselength}

To illustrate this,  we consider the simple triangular loop $\Graph(\Nodes=\ccup{1,2,3}, \Edges=\ccup{a,b,c})$ represented in Fig. \ref{fig:tree_toymodel_panel}, with $M=2$ commodities moving in the network and lengths $\ell = (\ell_{a},\ell_{b},\ell_{c})$.
For simplicity we focus on the phase diagram in $\ccup{\ell_{e}}$ by fixing $S$, but similar reasoning applies when doing the opposite.
We set $S^1=(+1,-1,0), S^2=(-1, +2, -1)$.
For this simple case, \kl \text{} law allows only for three possible tree topologies $\mathcal{T}_{i},\, i=1,2,3$, these are shown on the bottom right of Fig. \ref{fig:tree_toymodel_panel}. By solving \kl \text{} law, we can write all the fluxes as a function of $F_{a}=\bup{F_{a}^{1},F_{a}^{2}}$. Then, by choosing two arbitrary values of $F_a^1, F_a^2$  we propose a loopy solution $\Graph_L$ to compare against the trees, this is the leftmost bottom triangle in Fig. \ref{fig:tree_toymodel_panel}. We show that there are values of $0 <\Gamma < 1$ for which this loopy solution has lower transportation cost than any of the trees.  One can compute all the costs using Eq. (\ref{eqn:cost}) (see Appendix.~\ref{secAPX:toyM} for details)  and then find values of $\{\ell^*=(\ell^*_a,\ell^*_b,\ell^*_c), F_a^{*}, \Gamma^*\}$ for which
\be
\label{eq:mincost_toymodel}
J_\Gamma(\Graph_L;\ell^*, F_a^{*}, \Gamma^*) \leq \min \{J_\Gamma(\mathcal{T}_{i};\ell^*, F_a^{*}, \Gamma^*)  \, : \, i = 1,2,3\}
\ee
holds. To find an example solution, one could fix certain values for these parameters and then numerically solve Eq. (\ref{eq:mincost_toymodel}); for few simple cases this can also be done analytically. We show an example phase diagram obtained by varying $\ell$ in Fig.~\ref{fig:phaseD}, where we plot the values of $\Gamma_{crit}$ such that
for $\Gamma\geq \Gamma_{crit}$, the cost $J_\Gamma(\Graph_L)$ is optimal, i.e. we have a phase transition between trees to loopy optimal topologies. Notice that such values of $\Gamma_{crit}$ depend on the selected values of $(\ell,\, F_{a})$, and that optimal loopy solutions are not guaranteed to exist for any arbitrary configuration of these values. This can be numerically investigated using similarly reasoning as done for the case above. The important point here is that we could find \textit{at least} one setting of $(\ell,\, F_{a})$ for which we have loopy solutions in the non-trivial regime $0<\Gamma<1$. Similar arguments can be used to find phase diagrams in $S$ when fixing $\ell$, see Appendix~\ref{secAPX:toyM}, Fig.~\ref{figAPX:phaseDS}.

\subsection{Phase diagram tree--loops (lengths fixed)}
\label{ssed:interaction_colors}

To make more clear the consequences of the implicit interaction between different fluxes when imposing a shared conductivity and the optimization process is run,
we show results on a simple synthetic toy model where we vary the load of one color, while keeping the others fixed.

 Specifically,  we study the triangle topology of Fig.~\ref{fig:tree_toymodel_panel} and consider two different configurations of $S$. The first  has $S_1^1 = -S_3^1 = 1$ for the yellow commodity, i.e. one unit of mass of type $i=1$ is moving from node $1$ to node $3$, while  the purple commodity has $S_2^2=+2$, and $S_3^2 = S_1^2 = -1$, i.e. two units of mass of type $i=2$ are injected in node $2$ and are equally split between destination nodes $1$ and $3$.  This corresponds to the green triangle in the phase diagram of Fig. \ref{fig:panel_interaction_fluxes}a. The second configuration has the same sources and sinks for the yellow commodity, while the purple mass is doubled, i.e.  $S_2^2=+4$, and $S_3^2 = S_1^2 = -2$. This corresponds to the red triangle in Fig.~\ref{fig:panel_interaction_fluxes}a.  As we can see from Fig.~\ref{fig:panel_interaction_fluxes}b, both the optimal network topologies and the fluxes of individual colors differ in the two configurations.  The important point here is that the fluxes of the yellow change, even though its forcing $S^{1}$ does not change between the two configurations. This is a consequence of having distinct commodities sharing a common infrastructure:  acting solely on the purple mass impacts also the path taken by the yellow, and consequently the overall optimal network topology.

Additionally, the way the topology changes between these two configurations depends on the exponent $\Gamma$. In this simple scenario,  we can have either a tree or a loop at $\Gamma=0.6$ ($\Gamma < \Gamma_{crit} \simeq 0.77$) or $\Gamma=0.8$ ($\Gamma > \Gamma_{crit}$) respectively, see Fig.~\ref{fig:panel_interaction_fluxes}b. In particular, the case of $\Gamma = 0.8$ is a simple example of how the routing mechanism is responsible for the generation of loops in a multi-commodity setting.  Finally, if we were to consider two separate uni-commodity scenarios and solve the optimization for the two colors independently, we would have obtained a different result, as shown in Fig.~\ref{fig:panel_interaction_fluxes}c. In this case, the yellow remains the same in the two configurations, while the purple would simply double the amount of fluxes along edges, but the set of edges being used would stay the same.

In addition to the numerical analysis presented to study the generation of loops,  in Appendix \ref{secAPX:trimxia} we adapt to our ``colored'' case the proof of Proposition 2.1 in Xia \citep{xia2003optimal}, where it was demonstrated that one-commodity (i.e. ``colorless'' \kl \text{} law) optimal transport paths are trees. Here we show that for our model optimal networks may contain loops.

\section{Conclusions}
Although we have a rigorous theoretical understanding of the behavior of one-commodity flows  in networks, comparable theoretical insights for flows of different types have been lacking.
Here we propose a model for multi-commodity flows that extends and generalizes various results obtained for the one-commodity case.
It assumes that all the commodities have the same priority by imposing their conductivities to be equal and that their dynamics is regulated by the 2-norm squared of the fluxes.
By drawing from theoretical results of optimal transport theory, the equilibrium solutions of our dynamics are also stationary points of a cost function that can be interpreted as the sum of operating and infrastructural costs. As we tune a parameter $\beta$,  our dynamics  can solve various type of routing optimization problems. Its numerical implementation is efficient and scalable to large systems. \\
Remarkably, our model shows how optimal loopy topologies can arise from simple dynamical rules. We explain how this emerges as a consequence of the colored \kl \text{} law and how the theoretical proof valid in one-commodity fails when fluxes are vectors. We provide example phase diagrams  on a simple toy model that illustrates how optimal topologies evolve from being trees to containing loops.

Our model is applicable to all situations where it is relevant to distinguish flow types and to consider how these interact. One important example of such an instance is in communication networks where packets of information need to be delivered at different destinations.

In our formulation the underlying network topology is given in terms of sets of nodes and edges. While our model allows for edge removal (and node removal as a consequence), it does not provide a mechanism for adding new connections. In order to allow for this, the natural modification of our approach would be to consider a continuous formulation as in \cite{facca2020branch, baptista2020network}. In this case, we would have no underlying topology to start with, except the presence of sources and sink nodes at given locations in space. This is an interesting direction for future work.

In addition to solving multi-commodity problems, our model allows to draw a rigorous mapping between two different formalisms. In fact, while both the physics and optimal transport communities are actively investigating these systems, we still miss a clear connection between them, even for the one-commodity case. We make a first attempt to fill this gap by showing how our dynamics maps to a standard optimization setup while also generalizing to the multi-commodity case. Furthermore,  we deploy two numerical methods that have lower computational complexity compared to others based on gradient descent.

We expect that our formalism can be further extended in the future to accommodate for more sophisticated interaction between commodities or in multilayer networks \cite{ibrahim2021optimal} thus better representing specific application scenarios. Similarly, modifying the dependence of the fluxes in driving the dynamics and investigating possible mappings to suitable optimization setups are natural next steps.

We foresee that the insights gained about the structure of optimal topologies and in combining principles of optimal transport and physics will open the way to further studies targeting these systems.
To facilitate this, we provide an open source implementation of our code at \href{https://github.com/aleable/McOpt}{\texttt{https://github.com/aleable/McOpt}}.

\section*{Acknowledgements}
We thank Kurt Mehlhorn for useful discussions and for the helpful advice about the proof of the Lyapunov functional. We acknowledge the help of Daniela Leite for data pre-processing. The authors thank the International Max Planck Research School for Intelligent Systems (IMPRS-IS) for supporting Alessandro Lonardi.

\appendix

\section{The Lyapunov functional}
\subsection{The Lyapunov functional is well-defined}
\label{secAPX:lyapunov_dyn}
Here we prove that the functional proposed in Eq. (\ref{eqn:lyap}) is a Lyapunov for the dynamics in Eq. (\ref{eqn:dyn}) for $0<\beta<2$.  To do that, we follow the derivations proposed in \cite{bonifaci2020} for a similar problem. We need to show:  (i) $\mathcal{L}_\beta \geq 0$,  (ii)  $\dot{\mathcal{L}}_\beta \leq 0$ and $\dot{\mathcal{L}}_\beta = 0$ if and only if $\{\hat{\mu}_e\}$ is a stationary point for the dynamics. The first condition is trivial. In order to prove the second one, we first define the quantity:
\be
L_{vu}=\sum_e B_{ve}(\hat{\mu}_e/\ell_e)B_{ue}\quad,
\ee
which is the entry $(u,v)$ of the Laplacian of a graph with adjacency matrix with entries $A_{uv}=\hat{\mu}_{uv}/\ell_{uv}$. We can thus rewrite Eq.~(\ref{eqn:kirkoff})-(\ref{eqn:constr}) as:
\bea
\sum_{e,u} B_{ve}\frac{\hat{\mu}_e}{\ell_e} B_{ue} p_u^i &=& S_v^i \quad \forall v \in \Nodes,  \forall i = \CommSet \\ \label{eqn:kirch_derAPX}
\sum_u L_{vu} p_u^i &=& S_v^i \quad \forall v \in \Nodes, \forall i = \CommSet \quad.
\eea
Now, we claim that for each edge:
\bea
\partial_{\hat{\mu}_e}\mathcal{L}_\beta &=& \f{1}{2} \bup{ \ell_e \hat{\mu}_e^{1-\beta} + \partial_{\hat{\mu}_e} \left( \sum_{i,u} S_u^i p_u^i \right) } \\
\label{eqn:claimAPX}  &\overset{\text{claim}}{=}& \f{\ell_e}{2} \left(\hat{\mu}_e^{1-\beta}-\f{||\Delta P_e||_2^2}{\ell_e^2} \right) \quad.
\eea
This identity can be obtained differentiating for $\hat{\mu}_e$ both sides of Eq. (\ref{eqn:kirch_derAPX}). This yields for all $e, v, i$ the following:
\bea
\sum_u (\partial_{\hat{\mu}_e}L_{vu}) \, p_u^i &+& \sum_u L_{vu} (\partial_{\hat{\mu}_e} \, p_u^i) =0\\
\label{eqn:tempAPX}
 \sum_u L_{vu} (\partial_{\hat{\mu}_e} \, p_u^i)  &=&- \sum_u B_{ve}(1/\ell_e)B_{ue} \, p_u^i \quad.
\eea
 Multiplying both sides of Eq. (\ref{eqn:tempAPX}) by $p_v^i$, summing over $v$, and exploiting again Eq. (\ref{eqn:kirch_derAPX}) on the l.h.s. we obtain:
\bea
\sum_{v,u} p_v^i L_{vu} (\partial_{\hat{\mu}_e} \, p_u^i)  &=&- \sum_{v,u} p_v^i B_{ve}(1/\ell_e)B_{ue} \, p_u^i \\
\partial_{\hat{\mu}_e} \left( \sum_u  S_u^i p_u^i \right) &=& - \ell_e\f{(\Delta P_e^i)^2}{\ell_e^2}
\eea
with $\Delta P_{e}$ being an $M$-dimensional vector of entries $\Delta P_{e}^{i}=\Delta P_{uv}^{i}=p^{i}_{u}-p^{i}_{v}$, with $e=(u,v)$. Summing over $i$ gives:
\be \label{eqn:identity_lyap_proofAPX}
\partial_{\hat{\mu}_e} \left( \sum_{i,u} S_u^i p_u^i \right) = - \ell_e\f{||\Delta P_e||_2^2}{\ell_e^2} \quad,
\ee
notice that the term in parentheses in the l.h.s. is exactly the ``operating cost'' of Eq. (\ref{eqn:lyap}). Using Eq. (\ref{eqn:identity_lyap_proofAPX}) the claim in Eq. (\ref{eqn:claimAPX}) immediately follows. It is in force of Eq. (\ref{eqn:claimAPX}) that we see that the Lie derivative of $\mathcal{L}_\beta$ is not positive. Namely:
\bea
\dot{\mathcal{L}}_\beta &=& \sum_e (\partial_{\hat{\mu}_e}\mathcal{L}_\beta) \, \dot{\hat{\mu}}_e \\
&=& - \sum_e \f{\ell_e}{2} \hat{\mu}_e^\beta \left(\hat{\mu}_e^{1-\beta}-\f{||\Delta P_e||_2^2}{\ell_e^2} \right)^2 \leq 0
\eea
where $ \dot{\hat{\mu}}_e$ has been subsituted with the r.h.s of Eq. (\ref{eqn:dyn}). Moreover, $\dot{\mathcal{L}}_\beta=0$ if and only if $\hat{\mu}_\Iedge = 0$ or $\hat{\mu}_e^{3-\beta} = \hat{\mu}^2_e||\Delta P_e||^2_2/\ell^2_e = ||F_e||^2_2$. This exact condition can be recovered setting $\dot{\hat{\mu}}_e = 0$ in Eq. (\ref{eqn:dyn}) and exploiting Eq.  (\ref{eqn:flow}). In particular, for each $e$ edge we get:
\bea \label{eqn:scalingAPX}
 \hat{\mu}_{e}^{1-\beta}&=&\f{\sum_{i}\bup{p^{i}_{u}-p_{v}^{i}}^{2}}{\ell_{e}^2}=\sum_{i} \bup{\f{\ell_{e}	\,F_{e}^{i}}{\hat{\mu}_{e}}}^{2} \f{1}{\ell_{e}^{2}}\\
    \hat{\mu}_{e}^{3-\beta} &=&  ||F_{e}||^{2}_{2} \label{eqn:scalingdynAPX}\quad.
\eea
\subsection{Equivalence between the Lyapunov transportation cost and the dissipated energy}\label{secAPX:lyap_lagrange}
We prove that the transportation cost $J=\frac{1}{2}\sum_{e} \,\f{\ell_{e}}{ \hat{\mu}_{e}}\,||F_{e}||^{2}_{2}$ is indeed identical to the first term of the Lyapunov functional of Eq.~(\ref{eqn:lyap}).  In fact,  combining Eq. (\ref{eqn:kirkoff})-(\ref{eqn:constr}) we can rewrite Kirchhoff's law as
\be
\label{eqn:kirchAPX}
\sum_{e,u} B_{ve}\frac{\hat{\mu}_e}{\ell_e} B_{ue} p_u^i = S_v^i \quad \forall v \in \Nodes, \forall i = \CommSet \quad.
\ee
Multiplying both sides of the equation for $p_v^i$ and summing over $i$ and $v$ yields
\be
\sum_e \f{\ell_e}{\hat{\mu}_e} ||F_e||_2^2 = \sum_{i,v}p_v^i S_v^i \quad,
\ee
which is the equality we wanted to show.

\section{Mapping the dynamics to an optimization problem}\label{secAPX:dyn2opt}

We show that a constrained optimization problem with a cost function representing the total dissipated energy over the whole network has a solution with the same scaling as in Eq. (\ref{eqn:scaling}).

Formally, consider the constrained optimization problem of Eq.~(\ref{eqn:opt1})-(\ref{eqn:opt3}).
This can be turned into an unconstrained optimization problem by introducing Lagrange multipliers:
\bea
\begin{split}
\label{eqn:JBetaAPX}
J_\beta (\ccup{\hat{\mu}_e},\ccup{F_{e}}) \,= \,&\displaystyle\f{1}{2}\sum_{e}\f{\ell_{e}}{\hat{\mu}_e}\, ||F_{e}||^{2}_{2}\\
+\,\frac{\lambda}{2(2-\beta)}&\bup{\sum_{e}\ell_{e}\hat{\mu}_e ^{2-\beta}-K^{2-\beta}} \\
+\sum_{v,i} \chi_{v}^{i}&\bup{\sum_{e}B_{ve}\,F_{e}^{i}-S_{v}^{i}} \quad.
\end{split}
\eea
Here we introduced a multiplicative factor $1/2(2-\beta)$ for the Lagrange multiplier $\lambda$ to ease calculations. Taking the partial derivatives w.r.t. $\hat{\mu}_e $ and setting them to zero (the optimality condition on the derivative of $J_\beta$ with respect to $F_e$ will be treated later on) yields, for each edge,
\be\label{eqn:LscalingAPX}
\lambda \,\hat{\mu}_e^{3-\beta}= ||F_{e}||^{2}_{2} \quad \ra \quad \hat{\mu}_e = \f{1}{\lambda^{1/(3-\beta)}}\,||F_{e}||_{2}^{2/(3-\beta)} \, .
\ee
This is the same scaling relationship obtained from the stationary state of the dynamics in Eq. (\ref{eqn:scaling}), up to a multiplicative constant.  It is also the natural ``colored'' generalization of the one-commodity case presented in \cite{maritan, bohn2007structure, kirkegaard2020optimal}, where instead of having $||F_{e}||_{2}$ one has the absolute value $|F_{e}|$, as $F_{e}$ is a scalar quantity there.
Imposing the global constraint in Eq. (\ref{eqn:opt2}) allows to determine the value of the multiplier $\lambda$:
\be
\displaystyle\sum_{e}\ell_{e}\hat{\mu}_e^{2-\beta} = \sum_{e}\ell_{e}\f{||F_{e}||^{2(2-\beta)/(3-\beta)}_{2}}{\lambda^{(2-\beta)/(3-\beta)}} = K^{2-\beta} \, ,
\ee
yielding
\be
\lambda = \f{1}{K^{3-\beta}}\, \bup{\sum_{e}\ell_{e}\,||F_{e}||^{2(2-\beta)/(3-\beta)}_{2}}^{(3-\beta)/(2-\beta)} \,.
\ee
Substituting back into Eq. (\ref{eqn:LscalingAPX}) we obtain
\bea\label{eqn:muscalingAPX}
\hat{\mu}_e &=& \f{||F_{e}||^{2/(3-\beta)}_{2}}{ \bup{\sum_{e}\ell_{e}\,||F_{e}||^{2(2-\beta)/(3-\beta)}_{2}}^{1/(2-\beta)}}\, K \quad .
\eea
Setting $\gamma=2-\beta$, we get the scaling
\bea
\label{eqn:scalingbon}
\hat{\mu}_e &\sim& \bup{ ||F_{e}||_{2}^{2}}^{1/(1+\gamma)} \quad,
\eea
which is analogous to that of the one-commodity case Eq. (5) in \cite{bohn2007structure}. The same exact scaling can be recovered from our dynamics by setting $\dot{\hat{\mu}}_e = 0$ as shown in Eq. (\ref{eqn:scalingAPX})-(\ref{eqn:scalingdynAPX}).

The total dissipation is obtained by substituting Eq.~(\ref{eqn:muscalingAPX}) inside Eq. (\ref{eqn:opt1}), leading to:
\bea
J_{\beta} &=&  \f{1}{2K}\,  \bup{ \sum_{e}\ell_{e}\,||F_{e}||_{2}^{2(2-\beta)/(3-\beta)}}^{(3-\beta)/(2-\beta)} \\
&=& \f{1}{2K}\,   \bup{ \sum_{e}\ell_{e}\,||F_{e}||_{2}^{2 \gamma/(1+\gamma)}}^{(1+\gamma)/\gamma}   \quad.
\eea
This cost is again analogous to that of the one-commodity case: Eq. (6) in \cite{bohn2007structure} for $\gamma=2-\beta$. Using similar arguments, i.e. noticing that the function $x^{(3-\beta)/(2-\beta)}=x^{(1+\gamma)/\gamma}$ is monotonically increasing for $0<\gamma=2-\beta<2$, the original minimization problem reduces to that of minimizing with respect to $\ccup{F_{e}}$ the cost of Eq. (\ref{eqn:cost}):
\be\label{eqn:JGammaAPX}
J_{\Gamma}(\ccup{F_{e}}) = \sum_{e}\ell_{e}\,||F_{e}||_{2}^{2(2-\beta)/(3-\beta)}= \sum_{e}\ell_{e}\,||F_{e}||_{2}^{\Gamma} \quad,
\ee
where $\Gamma=2(2-\beta)/(3-\beta)=2\gamma/(1+\gamma)$, which is analogous to the model of Banavar et al. \cite{maritan}.
Lastly, we can set to zero also the derivative w.r.t. $F_{e}^{i}$ in Eq. (\ref{eqn:JBetaAPX}):
  \bea\label{eqn:fluxes_iterationAPX} \f{\partial J_\beta}{\partial
    F_{e}^{i}} &=& \f{\ell_{e}}{\hat{\mu}_{e}}\, F_{e}^{i}
  +B_{ue}\,\chi_{u}^{i}+B_{ve}\,\chi_{v}^{i}\\ \label{eqn:fluxes_iteration_2APX} &=&
  \f{\ell_{e}}{\hat{\mu}_{e}}\, F_{e}^{i}
  +B_{ue}\bup{\chi_{u}^{i}-\chi_{v}^{i}}\mbeq
  0\quad,\\\rightarrow F_{e}^{i} &=&
  -\f{\hat{\mu}_{e}}{\ell_{e}}\,B_{ue}\bup{\chi_{u}^{i}-\chi_{v}^{i}}
  \quad, \label{eqn:fluxes_iteration_3APX}\eea
  recovering the classical result stating that the
  pressure $p$ is (minus) the Lagrange multiplier obtained when we
  minimize the dissipated energy $J_{\beta}(\ccup{ \hat{\mu}_e},\ccup{F_{e}})$ under
  the Kirchhoff's laws constraints.


\section{Phase diagram for a toy model}\label{secAPX:toyM}

Here, we discuss in more detail the computations described in Sec.  \ref{sec:phaselength} to enforce the claim that networks with loops can be optimal for $0 < \Gamma < 1$.
The simple triangular loop of  Fig. \ref{fig:tree_toymodel_panel} (top) admits three possible tree topologies $\mathcal{T}_{i},\, i=1,2,3$,  drawn at the bottom right of Fig. \ref{fig:tree_toymodel_panel}.
Exploiting \kl \text{} law we write the fluxes as a function of $F_{a}=\bup{F_{a}^{1},F_{a}^{2}}$.  Then,  computing all the costs using Eq. (\ref{eqn:cost}), we get:
\begin{align}
J_\Gamma(\mathcal{T}_{1}) &= 2^{\Gamma}\, \ell_{b}+ 2^{\Gamma / 2}\, \ell_{c} \\
J_\Gamma(\mathcal{T}_{2})  &=2^{\Gamma}\, \ell_{a}+ 2^{\Gamma / 2}\, \ell_{c} \\
J_\Gamma(\mathcal{T}_{3}) &=\bup{\ell_{a}+\ell_{b}}\, 2^{\Gamma / 2}\\
\begin{split}
J_\Gamma(\Graph_L) &= ((F_a^1)^2 + (F_a^2)^2)^{\Gamma/2}\, \ell_{a}\\&+ ((F_a^1)^2 +(F_a^2 + 2)^2)^{\Gamma/2}\, \ell_{b}\\&+((F_a^1  -1)^2 + (F_a^2 + 1)^2)^{\Gamma/2}\, \ell_{c}\, \quad.
\end{split}
\end{align}
Thus, we need to find values of $\{\ell^*=(\ell^*_a,\ell^*_b,\ell^*_c), F_a^{*}, \Gamma^*\}$ for which Eq. (\ref{eq:mincost_toymodel}) is satisfied. In practice, the lengths are usually given in input, thus we
set $\ell_{a}=\ell_{b}= \f{3}{2}\,\ell_{c}, \,\ell_c = 1$; we then propose $F^*_a=(0,-1)$. Thus:
\bea
J_\Gamma(\mathcal{T}_{1})=J_\Gamma(\mathcal{T}_{2}) &= & \f{3}{2}\times 2^{\Gamma}+ 2^{\Gamma / 2}\label{eqn:toyMJ1APX} \\
J_\Gamma(\mathcal{T}_{3}) &= &3 \times 2^{\Gamma / 2}\label{eqn:toyMJ2APX}\\
J_\Gamma(\Graph_L) &= & 4 \label{eqn:toyMJ3APX}\quad.
\eea
Analytically from Eq. (\ref{eqn:toyMJ1APX})-(\ref{eqn:toyMJ3APX}) or numerically solving Eq. (\ref{eq:mincost_toymodel}) (fixing $F_a^*$ as proposed for $\Graph_L$) one can show that $\Gamma^* \simeq 0.83$.  Notice that such $\Gamma^*$ is not optimal, in the sense that for other choices of $F_a$ we may find lower values of the exponent $\Gamma$ enabling loopy networks to be optimal, we denote the minimum of this values as $\Gamma_{crit}$. Numerically solving Eq. (\ref{eq:mincost_toymodel}) for the toy model just discussed returns $\Gamma_{crit} \simeq 0.77$,  as shown in Fig. \ref{fig:phaseD} (green star). However, notice that the key point of this derivation is that we could find \textit{at least} one choice of $(\ell,\, F_{a})$ for which we have loopy solutions in the non-trivial regime $0<\Gamma<1$. Indeed, at  $\Gamma=\Gamma^*$ we have $J_\Gamma(\Graph_L;\Gamma)=J_\Gamma(\mathcal{T}_{1}) =J_\Gamma(\mathcal{T}_{2}) =J_\Gamma(\mathcal{T}_{3})=4$.

\begin{figure}[htp]
\begin{center}
 \includegraphics[width=0.9\columnwidth,trim={0cm 0.5cm 0cm 0.1cm},clip]{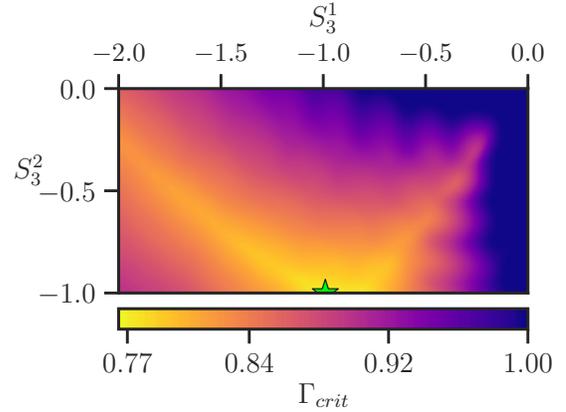}
\caption{\label{figAPX:phaseDS} Phase diagram in $S$. This can be numerically found by fixing $\{\ell_{e}\}$ as in Fig.~\ref{fig:tree_toymodel_panel}, $S_2^1 =  - 1 - S_3^1,\, S_1^2 = - 2 - S_3^2$, while varying $S_3^1$ and $S_3^2$. The colorbar denotes the value of $\Gamma_{crit}$ above which optimal solutions can be loopy. The green star denotes the configuration of $S$ used in Fig.~\ref{fig:tree_toymodel_panel}. }
\end{center}
\end{figure}

The same procedure can be used to find the phase diagram of Fig. \ref{figAPX:phaseDS}, here the costs $\{J(\mathcal{G}), J(\mathcal{T}_i)\}$ have been computed fixing the lengths as $\ell_{a}=\ell_{b}= \f{3}{2}\,\ell_{c}, \,\ell_c = 1$, and using $(F_a, S_3^1,S_3^2)$ as independent variables.

\section{Trimming loops to obtain trees}\label{secAPX:trimxia}
\begin{figure}[htpb!]
\begin{center}
\includegraphics[scale=0.82,trim={0cm 0.7cm 0cm 0.3cm},clip]{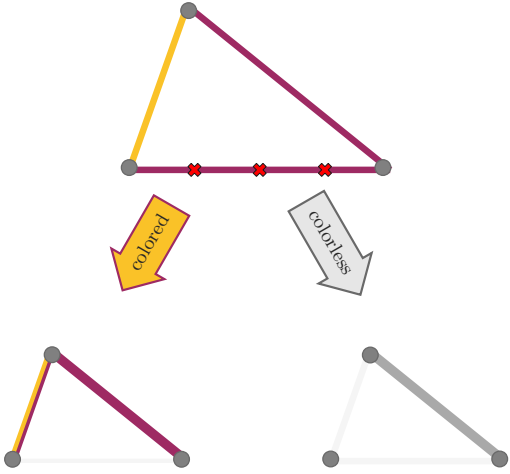}
\caption{\label{fig:tree_toymodel_trim} Sketch of the trimming procedure described in Appendix \ref{secAPX:trimxia}, on the toy model of Fig. \ref{fig:tree_toymodel_panel}. The tree obtained in the colored case is not optimal, while the one obtained in the colorless has lower cost but violates Kirchhoff's law. }
\end{center}
\end{figure}

Any configuration of the edge fluxes $\ccup{F_{e}}$ satisfying the ``colored'' \kl \text{}
law (\ref{eqn:kirkoff}) can be associated to a weighted graph
$\Graph(\Nodes,\Edges, W)$ with weights $w_e=\Ltwonorm{\Flux_e}$, where $\Nodes$ and $\Edges$ are the set of nodes and edges of the original input network. Denote with $\mathcal{T}$ the optimal tree
 topologies among these weighted graphs, i.e. loop-less topologies with weights minimizing $J_\Gamma$ as defined in Eq.~(\ref{eqn:cost}).
These trees (not necessarily unique) can be obtained by taking a weighted graph $\Graph_L$ with a single loop denoted as $L$, and cutting the loop by trimming  one of its edges, then redistributing the fluxes passing through the trimmed edge over the remaining links of $L$. We assign an arbitrary orientation $\hat{e}_\ell$ to the edges of $L$ so that $\langle \hat{e}_\ell, \hat{e} \rangle = \pm 1$, where the direction $\hat{e}$ of each link of a graph is uniquely determined by its incidence matrix. The edge to be cut is the one with smallest weight over the edges in the loop with a negative direction with respect to the graph's orientation.  Its flux is redistributed over the remaining edges, which now make a tree. Formally, we assign to the edges of $\mathcal{T}$ fluxes $F^{*}_{e}$ such that their entries are
 \be
 \label{eq:weight_loop_tree}
\bup{F_{e}^{i}}^{*} = F^\Icomm_\Iedge + \langle \hat{e}_\ell, \hat{e} \rangle\, F^\Icomm_{min} \quad \forall \Iedge \in \Edges, \forall \Icomm = \CommSet \quad,
 \ee
with $F_{min}=(F^{1}_{min},\dots,F^{M}_{min}) = \argmin_{e} \{\,\ell_e ||F_\Iedge||_{2}^{\Gamma} : \langle \hat{e}_\ell, \hat{e} \rangle =-1 \}$  and $F_{e}$ are the fluxes of $\Graph_L$. The orientation of $L$ can be switched in case the set of edges with negative orientation is empty. In the one-commodity case, as in Xia \citep{xia2003optimal}, there is a similar trimming, but with scalar weights on the tree being $F_{e}^{*}=F_{e}+ \langle \hat{e}_\ell, \hat{e} \rangle\,  F_{min}$, where now all the fluxes are numbers. The key effect of having a scalar trimming is that $F_{e}^{*}$ can become zero as a result of having a negative orientation $\langle \hat{e}_\ell, \hat{e} \rangle$; in words, the flux $F_{min}$ adds negatively to the fluxes originally present in $\Graph_L$ along the edges in the loop with negative orientation, if $F_{min}=F_{e}$ then $F_{e}^{*}=0$. Here instead, in Eq. (\ref{eq:weight_loop_tree}) we add a vector. While we might have that for certain components the flux cancels out, the norm of the whole vector $F_e$ might not be zero, because not all the components (colors) cancel. This results from imposing a colored \kl \text{} law.

We illustrate this procedure on the triangle network in Fig. \ref{fig:tree_toymodel_panel}. In particular,  Fig. \ref{fig:tree_toymodel_trim} shows how this trimming applies in the colored case (our case) against a colorless case. The tree obtained in the colored case is not optimal, while the one obtained in the colorless  has lower cost but is not valid, as it violates the constraints enforced by Kirchhoff's law. The consequence is that now loops can be the optimal solutions while in the one-commodity case optimal networks were trees.
Specifically, there exists a $\Gamma_{crit} \in (0,1)$ (or $\beta_{crit}\in (1,2)$) such that we have a phase transition between trees and loopy structures. The value of $\Gamma_{crit}$ depends on $\bup{S, \ccup{\ell_{e}}}$.

\section{Numerical implementation}\label{secAPX:NumImpl}

\subsection{Implementation details and gradient descent}\label{secAPX:algorithms}

We propose two approaches to solve our problem that are the natural multi-commodity generalization of those used in \cite{bohn2007structure, corson2010fluctuations, hu2013adaptation}. One is based on finding the steady state of the conductivities using Eq. (\ref{eqn:dyn}) (Dynamics), and one on implementing the iterative update of Eq. (\ref{eqn:iterativeMu}), and Eq. (\ref{eqn:iterativeF}) (Optimization). The implementations of these methods are summarized in Algorithms (\ref{dynamics_algAPX}) and (\ref{optimization_algAPX}).

\begin{algorithm}[H]
  \caption{Dynamics}
  \label{dynamics_algAPX}
   \begin{algorithmic}[1]
   \State{Input: $\Graph(\Nodes,\Edges) =$ adjacency list, nodes coordinates; $M$; inflows; $0 < \beta < 2$}
   \State{Initialize: (i) $S$ and (ii) $\{ \hat{\mu}_e \}$ (e.g. sampling as i.i.d. $\hat{\mu}_e \sim U(0,1)$)}
   \While{convergence not achieved}
   	\State{solve Kirchhoff's law as in Eq. (\ref{eqn:kirkoff}) $\rightarrow \{P_v^i\}$}
   	\State{update conductivities with a finite difference discretization of Eq. (\ref{eqn:dyn}): $\{\hat{\mu}^t_e\} \rightarrow \{\hat{\mu}^{t+1}_e\}$}
   \EndWhile
      \State{Return: fluxes $\{ F_e^i \}$ at convergence, computed using $F_e^i = {\hat{\mu}_e}(p^i_u - p^i_v)/\ell_e, \, e = (u,v)$}
   \end{algorithmic}
\end{algorithm}
\begin{algorithm}[H]
  \caption{Optimization}
  \label{optimization_algAPX}
   \begin{algorithmic}[1]
   \State{Input: $\Graph(\Nodes,\Edges) =$ adjacency list, nodes coordinates; $M$; inflows; $0 < \beta < 2$}
   \State{Initialize: (i) $S$ and (ii) $\{ \hat{\mu}_e \}$  (e.g. sampling as i.i.d. $\hat{\mu}_e \sim U(0,1)$)}
   \While{convergence not achieved}
   	\State{solve Kirchhoff's law as in Eq. (\ref{eqn:kirkoff}) $\rightarrow \{P_v^i\}$}
    \State{update fluxes using Eq. (\ref{eqn:iterativeF})}
   	\State{compute $\hat{\mu}_e(F_e)$ using Eq. (\ref{eqn:iterativeMu})}
   \EndWhile
      \State{Return: fluxes $\{ F_e^i \}$ at convergence}
   \end{algorithmic}
\end{algorithm}

 \begin{figure*}[htp]
\begin{center}\label{fig:complexity}
\subfloat[\label{subfig:aAPX}]{
  \includegraphics[scale=0.95,trim={0cm 0.3cm 0cm 0cm},clip]{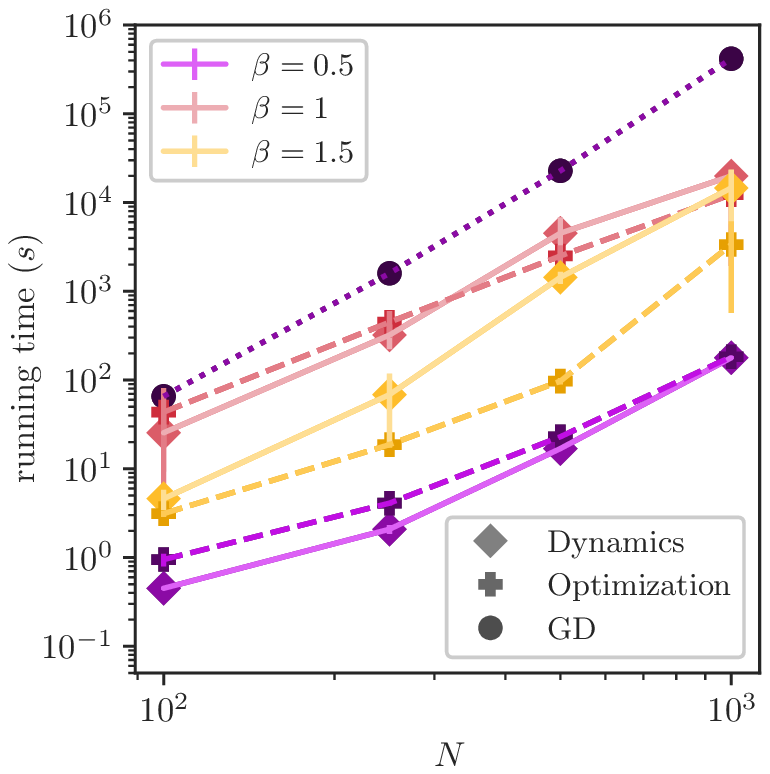}
}
\subfloat[\label{subfig:bAPX}]{
  \includegraphics[scale=0.95,trim={0cm 0.3cm 0cm 0cm},clip]{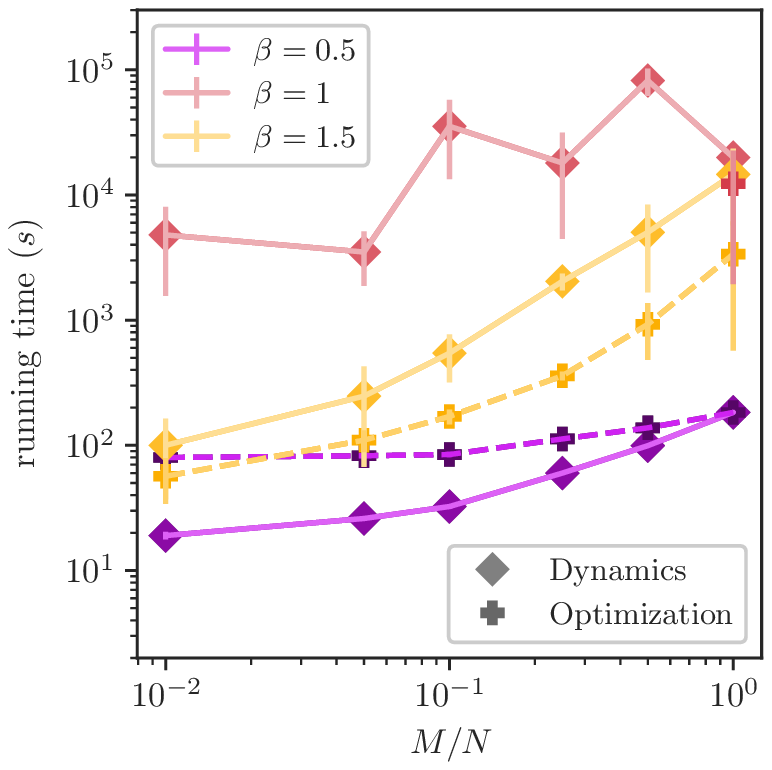}
}
\caption{Computational complexity. (a) Running time (in seconds) as a function of system size $N$. (b) Running time as a function of the ratio $M/N$ between the number of commodities and system size. GD denotes gradient descent, implemented with Eq. (\ref{eqn:gd1APX}) and (\ref{eqn:gd2APX}); we only show it for $\beta=0.5$ as for the other values it fails to converge within a reasonable time. Similarly, for $\beta=1$ and $M/N<1$, Optimization fails to converge, hence we only report Dynamics.}
\end{center}
\end{figure*}

These pseudo-codes outline our methods, however practitioners can make further arbitrary choices about what numerical routines to use in the various steps. In our implementation, we solved $M$ ordinary differential equations as in Eq. (\ref{eqn:dyn}) by means of an explicit Euler method, thus at each step the local truncation error is approximately proportional to $\Delta t^{2}$, with $\Delta t$ being the difference between two consecutive time steps, which can be arbitrarily set in input. Solutions of \kl \text{} law have been computed using a sparse direct solver.\\ Lastly, we impose the following convergence criteria: \textit{convergence} is achieved when these conditions are satisfied
\bea
\text{Dynamics: }\max_e|\hat{\mu}_e^{t+1} - \hat{\mu}_e^t|/\Delta t < \tau_{dyn} \quad, \\ \text{Optimization: }\max_e\left| ||F_e||_{2}^{t+1} - ||F_e||_{2}^t \right| < \tau_{opt}
\eea
where $\tau_{dyn},\tau_{opt} > 0$ are parameters arbitrarily set in input. In our experiments we use $\tau_{dyn}=10^{-3}$, $\tau_{opt}=10^{-5}$. To test our methods, we  developed a momentum-based gradient descent as baseline algorithm. This consists in the component-wise iterative update of the fluxes using
\bea
\bup{V_e^i}^t &=& \eta \, (\partial J_\beta/ \partial \Flux_\Iedge^i)^t + \delta\, \bup{V_e^i}^{t-1} \label{eqn:gd1APX}\\
({\Flux_\Iedge^i})^{t+1} &=& (\Flux_\Iedge^i)^{t} + (V_\Iedge^i)^{t} \label{eqn:gd2APX}
\eea
with $\eta, \delta > 0$ fixed increment rates  and $\bup{V_{e}^{i}}^{0}=F_{e}^{i}$. We fixed the convergence criteria analogously to what done for the other two methods: $\max_e\left| ||F_e||_{2}^{t+1} - ||F_e||_{2}^t \right|/\eta < \tau_{gd}$, with $\tau_{gd} > 0$ a parameter that needs to be set in input. In our experiments we set it to $\tau_{gd} = 10^{-2}$.
From a theoretical point of view, the comparison with a standard gradient descent method was proposed in light of the equivalence of our dynamics and a mirror-descent approach for
the Lyapunov functional, as proved for $\beta=1$ in
\cite{bonifaci2019laplacian}. The dynamics automatically
preserves positiveness of the conductivities $\{ \hat{\mu}_\Iedge \}$, thus a large time-step
can be used. On the contrary, using purely gradient descent approaches, the time step
size must be reduced when some entries of the vector $\{ \hat{\mu}_\Iedge \}$ go to zero.
After running our algorithms until convergence, the original network is trimmed by removing edges with negligible fluxes. Formally, we remove links for which $||F_e||_2 < \tau$, with $\tau>0$ arbitrarily fixed. Typically, as we empirically found, the distribution of $||F_e||_2$ over the edges is divided in two sets having values differing by several orders of magnitude. It is thus straightforward to distinguish what edges to be trimmed, i.e. those that have negligible values compared to the rest of the distribution.

\subsection{Computational complexity}\label{secAPX:compcost}

Each temporal step executed in our algorithms requires the approximate solution of $M$ linear systems of dimension $N$. This operation has been carried out by means of a sparse direct solver (UMFPACK) that performs a LU decomposition for each column of the right hand side of Eq. (\ref{eqn:kirkoff}). The total computational complexity of this process scales as $\mathcal{O}(MN^2)$. To have a better understanding of this we tested our models with several synthetic Waxman networks obtained by placing $N$ nodes uniformly at random in a square domain of size $1$. Nodes  are connected with probability $p = A \,\exp\bup{-d / \alpha L}$, where $A,\alpha,L$ are parameters that we fix arbitrarily to $A = 1/4$, $\alpha = 1/4$ and $L=1$; $d$ is the euclidean distance between a pair of nodes. The matrix $S$ is constructed assigning a total inflowing mass of $10^4$ at random to $M$ nodes, and redistributing on the nodes of the network proportionally to their inflows.

We test the efficiency of our schemes by measuring the total running time (in seconds) to reach convergence for different values of $\beta$, $M,N$. Results are shown in Fig. \ref{fig:complexity}. We notice that the two algorithms Dynamics and Optimization have similar computational complexity. Their small running time's differences are negligible and only due to how convergence is precisely defined, i.e. how the corresponding parameters $\tau_{dyn}$, $\tau_{opt}$ are set. The running time is shorter for $\beta<1$ (traffic optimization, loopy) than in the opposite scenario of $\beta>1$ (minimization of infrastructural cost, tree-like). The case $\beta=1$ is more nuanced, as the cost transitions between two opposite situations. In this case,  Optimization fails to converge for $M/N<1$, if convergence is defined in terms of variations of $||F_{e}||_{2}$ between iteration steps. This is because the algorithm gets lost in degenerate local minima, configurations with same cost but different set of fluxes. This lack of convergence suggests that, for $\beta=1$, the energy landscape around these minima is flat, i.e. there are many configurations with same cost but non-negligible differences in their fluxes. The Optimization routine keeps switching between these different states. In this case, one can simply pick one of these possible many solution as an example local optima. The dynamics does instead converge. This suggests that Dynamics is biased towards one of these degenerate solutions. For $M/N=1$, Optimization converges with same running time as Dynamics, suggesting that as we enlarge $M$ the landscape becomes less flat. A possible cause is that by increasing $M$ the system has more constraints to be satisfied via \kl \text{} law, which reduces the number of possible degenerate solutions. This claim is also supported by the behavior of Dynamics' running time, which does not monotonically increase with $M/N$ in this case, as shown in Fig. \ref{fig:complexity}b. These behaviors highlight relevant differences between the two implementations. Finally, we note that the computational complexity could in principle be further reduced to $\mathcal{O}(M N)$ using multigrid methods \cite{briggs2000multigrid}, we do not explore this here.

\bibliography{bibliography}

\end{document}